# Empathy on Demand: How Empathic AI Can Scale Emotional Support for Verbal Harassment

Anouk Bergner*

University of Geneva (anouk.bergner@unige.ch)

Geneva, Switzerland

Philipp Winder

University of St. Gallen (philipp.winder@unisg.ch)

St. Gallen, Switzerland

Christian Hildebrand

University of St. Gallen (christian.hildebrand@unisg.ch)

St. Gallen, Switzerland

**Abstract**

Verbal harassment is a growing source of psychological stress for people around the world. It occurs both online and offline and relies on language to demean, threaten, or discredit its targets. Unlike other stressors such as loss or uncertainty, verbal harassment aims at silencing its targets by eroding their sense of being heard and weakening their perceived ability to respond. Many individuals lack access to adequate and timely support, however, when they experience such harassment. People increasingly turn to conversational artificial intelligence (AI) such as ChatGPT or dedicated AI companions for emotional support, raising questions about whether it can facilitate the same psychological benefits as actual human empathy. We focus on online contexts as a prevalent application of verbal harassment. We develop and test a psychological framework identifying three key linguistic signals of empathic listening (perspective-taking, emotional validation, and action orientation), that together restore a sense of feeling heard and enhance coping in the context of verbal harassment. We find that LLMs consistently produce language exhibiting stronger empathic-listening markers than human non-experts and trained mental health professionals, promoting more approach-oriented (vs. avoidance-oriented) coping strategies. A subsequent behavioral study shows that these linguistic signals boost recipients' sense of feeling heard and increase their coping self-efficacy. These findings reveal how specific linguistic features create empathic connections between humans and advanced conversational AI and can enhance people's psychological resilience. Our results highlight the potential for AI to serve as a scalable source of emotional support, especially when human support is unavailable or insufficient.

## Introduction

Digital platforms are transforming how people interact socially. They offer new opportunities to connect and communicate, but they also create the potential to experience harm. Verbal harassment, which includes but is not limited to its online form, is a growing and urgent social challenge. It occurs when language is used to demean, threaten, or discredit others, and it can take place both in digital and offline contexts. Recent estimates indicate that approximately one in four adult internet users reported experiencing harassment[1], illustrating the scale of this issue.

Verbal harassment differs from other psychological stressors because its distress does not stem from contextual circumstances such as loss or uncertainty but from targeted efforts to silence an individual[2]. Such experiences undermine people's sense of being heard and their perceived ability to respond to the threat[3,4]. While news headlines often highlight extreme cases[5,6], even milder forms of verbal harassment can compromise individuals' coping resources and decrease overall well-being over time[7]. Despite growing awareness and an effort to expand traditional support mechanisms, existing resources often face inherent constraints. Professional mental health care can be costly, difficult to access, or perceived as disproportionate to the problem[8]. Peer support networks, though more accessible, can be inconsistent and emotionally risky[9]. Additionally, many victims of verbal harassment hesitate to seek support, anticipating stigma, negative judgment, or dismissal, leading to a growing support gap[10].

Conversational artificial intelligence is now increasingly stepping into this gap. Large language models (LLMs) widely accessible through tools like ChatGPT (OpenAI), Gemini (Google), or Claude (Anthropic) are frequently used not only for productivity or information retrieval but also for coping support[11,12]. For example, journalists targeted by online harassment now turn to Claude to receive empathic coaching and practical guidance on how to respond[13] and teens report using AI companions such as ChatGPT for emotional support and advice in personal or stressful situations[14]. LLMs growing capabilities in understanding and responding to human needs[9,15–19], together with their promise of private, nonjudgmental support, offer users a low-barrier, easy-access alternative to traditional forms of support, particularly in situations that feel too personal for public disclosure or too minor for clinical intervention.

In this research, we demonstrate that LLMs can provide effective empathic support that enhances psychological coping. We treat empathic listening as a linguistic construct and evaluate linguistic markers rather than experiential listening processes[20]. We find that LLMs consistently produce language exhibiting stronger empathic-listening markers than untrained humans and mental health professionals, leading support recipients to feel more heard and report greater coping self-efficacy. We do not propose that LLMs should replace therapeutic relationships or human social support. Instead, our findings point to their potential as accessible first responders or complementary resources that can address critical gaps in support provision and accessibility at a global scale.

**Theoretical Background**

Our theoretical approach centers on understanding how language conveys empathy in ways that promote psychological coping. We focus specifically on the experience of feeling heard, which is defined as the perception that one's perspective, emotions, and agency are recognized and respected[21]. Feeling heard may be particularly important in the context of verbal harassment, which often aims to silence or discredit individuals through language. Feeling heard is increasingly recognized as a key psychological process that promotes emotional recovery and social resilience[22–26], as it creates a psychological foundation for greater coping self-efficacy[27,28]. Enhanced coping self-efficacy, in turn, has been consistently linked to improved psychological well-being, including reduced stress, lower anxiety, and greater resilience[29–31].

While empathy has long been considered a multi-faceted construct[32,33], its communication in digital settings presents unique challenges. Prior research on empathic agents has focused mainly on modelling affective and behavioral expressions of empathy[34,35]. In text-based communication, however, support recipients need to rely mainly on semantic signals embedded in language to judge whether they are being genuinely heard and understood[36]. Thus, in the present work we treat empathic listening as a linguistic construct rather than an experiential process.

We propose that three specific signals of empathic listening create the subjective experience of feeling heard: perspective-taking, which conveys contextual understanding[37]; emotional validation, which validates an emotional experience[38,39]; and action orientation, which affirms an individual's agency[28]. Together, these elements represent distinct but complementary dimensions of empathic listening that map onto key psychological needs in support interactions[40,41]: the need to be understood (perspective-taking), feeling emotionally validated (emotional validation), and empowered to take back control (action orientation). We hypothesize that together, these linguistic signals of empathy create a context in which individuals feel genuinely heard by the support provider.

Empathic support can be valuable across many situations, but its importance will likely increase in more severe harassment situations where the psychological toll is greater and the need to restore agency is more urgent. In such situations, people benefit disproportionately from responses that affirm their agency and encourage constructive action to regain control[42]. Yet human supporters are more likely to withdraw emotionally due to compassion fatigue[43], emotional self-protection[44], or anxiety-related decreases in perspective-taking[45], creating a disconnect between what support seekers need and what (human) responders can provide. We therefore hypothesize that LLMs, which are not subject to emotional fatigue or self-protective distancing, may be particularly beneficial in more severe verbal harassment situations where consistent, agency-supporting guidance is most needed but human capacity is more likely to be compromised.

Taken together, we propose that empathic listening operates through three key linguistic features (perspective-taking, emotional validation, and action orientation) that collectively create the psychological experience of feeling heard. This experience, in turn, enhances individuals' coping self-efficacy when facing verbal harassment. We predict that LLMs will produce language with stronger empathic-listening signals compared to human supporters, particularly in more severe harassment situations.

**Overview of Studies and Methodological Pipeline**

We adopt a two-sided approach to understand how empathic language supports coping. On one side, we examine how empathic listening (perspective-taking, emotional validation, and action orientation) is encoded in support providers' language (Study 1), and on the other side, we assess their impact on support recipients' sense of feeling heard and ability to cope with a stressful situation (Study 2).

Consider, for example, two scenarios of real harassment cases in a social media context: in a mild case, a user received the message "You're an idiot." In a severe case, a user was confronted with: "You deserve to die." How would different types of supporters respond to help someone cope with such verbal harassment expressed in online communication? To answer this question, Study 1 compares how LLMs, laypeople, and trained mental health professionals respond to these online harassment scenarios. We ask each support provider how they would advise someone facing such harassment to respond. We then analyze these responses for linguistic empathy features and coping orientation. Table 1 provides examples of responses across provider types and harassment severity.

Table 1 here

In Study 2, we shift to the recipient perspective. A new panel of participants reads one of the harassment scenarios and a supportive response from Study 1. Participants rate how the response makes them feel and whether it helps them cope with the situation. This two-sided research design (Figure 1) enables us to assess not only the linguistic features of empathy in support providers' responses but also analyze their downstream psychological consequences on support recipients. Together, these two studies allow us to test our broader framework of how empathic language restores a sense of voice and agency after silencing in verbal harassment.

Figure 1 here

To implement this two-sided framework, we developed a multi-method research pipeline (Figure 2). First, we constructed a set of standardized verbal harassment prompts pretested for severity and linguistic comparability. Next, we collected 3,652 supportive responses from three provider groups: human non-experts, licensed mental health professionals, and four state-of-the-art, openly accessible LLMs. We analyzed each response using (1) zero-shot LLM-based ratings of empathic-listening features, (2) text-analytic assessment of linguistic style, and (3) classification of coping strategies. Finally, human participants evaluated a representative subset of these responses on perceived empathy, feeling heard, and coping self-efficacy.

All data and code are publicly available on OSF and all studies were preregistered and approved by the University of St. Gallen Institutional Review Board (Ethics Committee, reference HSGEC-20231218).

Figure 2 here

## Results

### Study 1: Comparative analysis of empathic listening features

We first compared how empathic signals are encoded in supportive responses from LLMs, human experts, and human non-experts. We used a zero-shot prompt with GPT-4o to rate perspective-taking, emotional validation, and action orientation (1 = very low to 7 = very high) for each of the 3,652 responses we collected. We also prompted GPT-4o to extract linguistic features related to each construct to validate the ratings. The combination of numerical rating and extracted features allowed us to assess whether GPT-4o correctly identified language that demonstrates understanding of each feature.

We find that LLMs consistently produce language exhibiting stronger empathic-listening markers than both human non-experts and human experts across all three features. For perspective-taking, LLM responses (M = 5.86, SD = 0.38) were rated higher than non-experts (M = 2.33, SD = 1.42; $t(625) = 60.70$, 95% CI [3.41, 3.64], $d = 4.91$, $p < .001$) and experts (M = 2.64, SD = 1.56; $t(668) = 52.05$, 95% CI [3.10, 3.34], $d = 4.05$, $p < .001$). For emotional validation, LLMs (M = 5.92, SD = 0.35) exceeded non-experts (M = 2.60, SD = 1.49; $t(620) = 54.49$, 95% CI [3.20, 3.44], $d = 4.51$, $p < .001$) and experts (M =

2.44, SD = 1.47; $t(667) = 59.78$, 95% CI [3.36, 3.59], $d = 4.67$, $p < .001$), while the difference between human experts and non-experts was non-significant ($t(1242) = -1.89$, $p = .059$). Mean differences were largest for action orientation, where LLMs (M = 6.82, SD = 0.45) outperformed both experts (M = 2.74, SD = 1.69; $t(672) = 60.86$, 95% CI [3.95, 4.21], $d = 4.67$, $p < .001$) and non-experts (M = 3.37, SD = 1.80; $t(622) = 46.69$, 95% CI [3.30, 3.60], $d = 3.83$, $p < .001$). Expert–non-expert comparisons were small or non-significant across the three markers, suggesting that, in brief text responses to decontextualized harassment prompts, experts did not systematically outperform laypeople on these linguistic dimensions.

We also find that responses systematically exhibited stronger empathic-listening markers in more severe harassment situations across all provider types. For perspective-taking, non-experts (mild: M = 2.13, SD = 1.32; severe: M = 2.53, SD = 1.48; $t(597) = -3.53$, 95% CI [-0.63, -0.18], $d = -0.29$, $p < .001$), experts (mild: M = 2.39, SD = 1.43; severe: M = 2.89, SD = 1.65, $t(633) = -4.18$, 95% CI [-0.75, -0.27], $d = -0.33$, $p < .001$), and LLMs (mild: M = 5.78, SD = 0.42; severe: M = 5.95, SD = 0.32; $t(2398) = -11.16$, 95% CI [-0.20, -0.14], $d = -0.46$, $p < .001$) all scored higher in severe compared to mild cases. A similar pattern emerged for emotional validation: non-experts (mild: M = 2.38, SD = 1.42; severe: M = 2.81, SD = 1.52; $t(601) = -3.60$, 95% CI [-0.66, -0.20], $d = -0.29$, $p < .001$), experts (mild: M = 2.21, SD = 1.31; severe: M = 2.67, SD = 1.58; $t(623) = -4.11$, 95% CI [-0.69, -0.22], $d = -0.32$, $p < .001$), and LLMs (mild: M = 5.88, SD = 0.37; severe: M = 5.96, SD = 0.33; $t(2366) = -5.90$, 95% CI [-0.11, -0.06], $d = -0.24$, $p < .001$). We found that action orientation had the strongest severity effects: non-experts (mild: M = 2.99, SD = 1.68; severe: M = 3.73, SD = 1.85; $t(599) = -5.15$, 95% CI [-1.02, -0.46], $d = -0.42$, $p < .001$), experts (mild: M = 2.43, SD = 1.54; severe: M = 3.04, SD = 1.78; $t(632) = -4.66$, 95% CI [-0.87, -0.35], $d$

= -0.37, $p < .001$), and LLMs (mild: M = 6.68, SD = 0.52; severe: M = 6.95, SD = 0.31; $t(2398) = -15.50$, 95% CI [-0.30, -0.24], $d = -0.63$, $p < .001$).

As can be seen in Figure 3, LLM scores were overall more tightly clustered across all three features, suggesting a higher and more consistent baseline of linguistically empathic communication. Results were identical when using non-parametric tests to allow for unequal variance, with all LLM-human and all severity contrasts remaining highly significant (all *p*s < .001). We also reran all analyses including response length (word count) as a covariate, and the results remained unchanged, indicating that differences in empathic language cannot be attributed to longer responses. These additional results are provided in the Supplementary Information on OSF (https://osf.io/hnzba/files/e9d6t?view_only=defc9f54efc04e5a811a58bae22a1b5c).

Figure 3 here

To shed further light on our observations, we conducted a qualitative analysis of the linguistic features of our support responses using Linguistic Inquiry and Word Count[46]. This analysis supports our finding that LLM responses were more consistent in the quality of their empathic features. Specifically, we find that LLM responses had higher Analytical Thinking scores (M = 35.83, SD = 0.57) compared to both non-expert humans (M = 21.75, SD = 1.14; $t(1849) = 9.41$, 95% CI [10.57, 17.60], $d = 0.54$, $p < .001$) and mental health professionals (M = 25.60, SD = 1.25; $t(1849) = 6.95$, 95% CI [6.78, 13.70], $d = 0.39$, $p < .001$). While not a direct measure of perspective-taking, analytical thinking facilitates a clear articulation and analysis of situational details, supporting perspective-taking[46]. Furthermore LLM-generated responses scored higher on Authenticity (M = 46.20, SD = 0.91) than both human experts (M = 36.21, SD = 1.46; $t(1849) = 5.52$, 95% CI [5.74, 14.20], $d = 0.31$, $p < .001$) and human non-experts (M = 28.67, SD = 1.38, $t(1849) = 9.52$, 95% CI [13.21, 21.90], $d = 0.55$, $p < .001$), signaling emotional resonance[47]. We also find that LLM-generated responses displayed a more positive emotional tone overall (M = 76.67, SD = 0.94) compared to both non-expert humans (M = 25.58, SD = 1.22; $t(1849) = 32.10$, 95% CI [47.40, 54.80], $d = 1.85$, $p < .001$) and mental health professionals (M = 26.69, SD = 1.14; $t(1849) = 32.00$, 95% CI [46.30, 53.60], $d = 1.81$, $p < .001$), reflecting LLMs' tendency to incorporate more positive emotional validation. Finally, we also find that LLM-generated responses scored higher on Clout (M = 76.10, SD = 0.67) compared to both human experts (M = 63.43, SD = 1.55; $t(1849) = 6.670$, 95% CI [8.21, 17.10], $d = 0.38$, $p < .001$) and human non-experts (M = 55.21, SD = 1.58; $t(1849) = 10.81$, 95% CI [16.36, 25.40], $d = 0.62$, $p < .001$), supporting the action-oriented nature of their responses[48].

Next, we examined both the number and the type of coping strategies suggested by each support provider. We first used GPT-4o to extract all distinct coping strategies mentioned in each response and structured the output as a list of individual strategies. Next, we applied keyword pattern matching to classify each strategy as either approach-oriented (e.g., confronting the harasser, seeking social support, setting boundaries) or avoidance-oriented (e.g., ignoring, blocking, emotional distancing), following established definitions in coping literature[49,50].

As shown in Figure 4, we find that LLMs provided significantly more coping strategies per response (M = 5.6, SD = 1.32) compared to human supporters (M = 1.30, SD = 1.14, $t(3650) = 98.12$, $d = 3.42$, $p < .001$).

Figure 4 here

These differences are not only quantitative, but also qualitative, as we find a distinct pattern in strategy recommendations. Specifically, as shown in Figure 5, LLMs offer a broader range of distinct coping strategies and a higher proportion of approach-oriented recommendations, such as confronting or reporting the harasser, seeking social support, or even using humor to respond. Human non-experts, in contrast, propose significantly more avoidance-oriented strategies, particularly distancing oneself from the situation and ignoring or blocking the harasser.

Overall, we find that human non-experts suggested avoidance in 75% of all strategies and approach in 26%, whereas human experts showed a more balanced distribution of strategies with 56% avoidance-focused and 44% approach-focused. LLMs were the most biased toward action, with 65% approach-focused versus 35% avoidance-focused strategies. These percentages reflect the relative frequency of strategy instances rather than the proportion of responses containing a particular strategy type, as many responses include multiple strategies. Approximately 5% of strategies could not be classified. This qualitative pattern reinforces our earlier findings that LLMs produce language that is more action-oriented than human support providers.

Figure 5 here

Taken together, our findings suggest that LLMs not only provide more empathic but also qualitatively different support compared to human supporters, with a consistent emphasis on more diverse and active coping strategies.

**Study 2: Recipient perceptions and psychological outcomes**

Next, we examined whether the differences in empathic language and coping strategies observed in Study 1 translate into differences in support recipients' perceptions and psychological outcomes. We recruited a new panel of participants (N = 431) to evaluate a representative subsample of support responses we collected in Study 1. To ensure both experimental control and high ecological validity, we implemented a multi-level randomization approach to assign participants to a verbal harassment scenario and sample a representative response from each support provider (GPT-4o, human non-experts, and mental health professionals, see Methods for additional details). After reading through each scenario case and the corresponding response, participants rated the response on the extent of feeling heard[21] and perceived coping self-efficacy[51]. To validate our three-feature model of linguistic empathy, participants also rated each response on perspective-taking, emotional validation, and action orientation, based on established scales. Finally, we included additional measures to explore the broader implications of feeling heard on participants' psychological well-being, the perceived helpfulness of the response, and to rank their overall response preference. These exploratory measures provide additional context for understanding how empathic listening impacts recipients beyond immediate coping.

Confirming our computational analysis from Study 1, we find that participants consistently rated LLM-generated responses as higher on all three empathic listening features. Specifically, as shown in Figure 6, LLM responses (M = 4.95, SD = 1.25) scored higher on perspective-taking compared to both human experts (M = 4.23, SD = 1.45, $t(860) = 7.75$, 95% CI [0.53, 0.90], $d = 0.53$, $p < .001$) and non-experts (M= 4.11, SD = 1.45, $t(860) = 9.16$, 95% CI [0.66, 1.02], $d = 0.62$, $p < .001$). For emotional validation, LLMs (M = 5.25, SD = 1.27) also scored higher than both human experts (M = 4.07, SD = 1.65, $t(860) = 11.81$, 95% CI [0.99, 1.38], $d = 0.80$, $p < .001$) and non-experts (M = 4.38, SD = 1.65, $t(860) = 8.68$, 95% CI [0.67, 1.06], $d = 0.59$, $p < .001$). Finally, we found the largest effect sizes for action orientation, where LLMs (M = 5.22, SD = 1.40) scored higher than experts (M = 3.61, SD = 1.77, $t(860) = 14.79$, 95% CI [1.40, 1.82], $d = 1.01$, $p < .001$) and non-experts (M = 3.67, SD = 1.89, $t(860) = 13.60$, 95% CI [1.32, 1.77], $d = 0.93$, $p < .001$), suggesting that LLM-generated responses most strongly outperformed human responses. Human non-experts and human experts did not differ significantly on perspective-taking ($p = .36$) and action orientation ($p = .83$), but experts scored lower on emotional validation ($t(860) = 2.82$, 95%

CI [0.10, 0.54], $d = 0.19$, $p < .01$), which may reflect professional communication norms emphasizing cognitive understanding and problem-solving over explicit emotional validation[52].

Figure 6 here

More importantly, greater perceived empathy translated into meaningful psychological outcomes. As shown in Figure 7, participants felt significantly more heard after receiving LLM responses (M = 5.43, SD = 1.28) compared to responses from human experts (M = 4.15, SD = 1.80, $t(860) = 12.04$, 95% CI [1.07, 1.49], $d = 0.82$, $p < .001$) and non-experts (M = 4.39, SD = 1.81, $t(860) = 9.76$, 95% CI [0.83, 1.25], $d = 0.66$, $p < .001$). We also find that participants felt significantly more able to cope with the situation after reading responses from LLMs (M = 5.45, SD = 1.14), compared to both human expert responses (M = 3.83, SD = 1.64, $t(860) = 16.90$, 95% CI [1.44, 1.81], $d = 1.15$, $p < .001$) as well as human non-expert responses (M = 3.96, SD = 1.53, $t(860) = 16.21$, 95% CI [1.31, 1.67], $d = 1.10$, $p < .001$), while the difference between experts and non-experts was non-significant ($p = .37$). We find that these effects even extended beyond coping self-efficacy to more positive overall well-being, with participants feeling greater overall well-being after reading responses from LLMs (M = 5.43, SD = 1.17) compared to responses from human experts (M = 3.99, SD = 1.69, $t(860) = 14.53$, 95% CI [1.25, 1.63], $d = 0.99$, $p < .001$) and human non-experts (M = 4.43, SD = 1.63, $t(860) = 10.33$, 95% CI [0.81, 1.19], $d = 0.70$, $p < .001$). The decrease in well-being following expert responses ($t(860) = 3.88$, 95% CI [0.21, 0.66], $d = 0.26$, $p < .001$) might reflect differences in supportive tone consistent with earlier patterns.

Figure 7 here

A moderated mediation model using orthogonal contrasts and 10,000 bootstrapped samples showed that GPT-4o responses, compared to human supporters (experts and non-experts), significantly increased perceptions of feeling heard, which in turn predicted greater coping self-efficacy. This indirect effect was stronger in severe ($\beta_{\text{GPT-Human}}$ = 0.57, 95% CI [0.46, 0.66]) than in mild harassment cases ($\beta_{\text{GPT-Human}}$ = 0.35, 95% CI [0.25, 0.45]; $\Delta\beta$ = 0.22, $p$ = .002). In contrast, the indirect effect comparing experts and non-experts was non-significant in mild cases ($\beta_{\text{Expert-Nonexpert}}$ = -0.04, 95% CI [-0.14, 0.06]) and only marginal in severe ones ($\beta_{\text{Expert-Nonexpert}}$ = -0.10, 95% CI [-0.20, -0.00]; $\Delta\beta$ = -0.07, $p$ = .35).

To ensure the robustness of our findings, we conducted several additional analyses. First, as LLM-generated responses were on average longer than human-generated responses, we controlled for response length in all models. This adjustment did not alter the significance or direction of any key effects, and response length was not a significant predictor in any model (all *p*s > .32), indicating that differences in perceived empathy and psychological outcomes could not be attributed to longer responses. Second, we replicated our primary analyses in a subsample of participants who reported having personally experienced verbal harassment in online contexts in the past (N = 239). Results in this subgroup closely mirrored the main findings, with LLM-generated responses outperforming both human experts and non-experts across all empathy dimensions and psychological outcomes (all *p*s < .001). This suggests that the observed effects are not limited to hypothetical evaluations but also hold among those with direct personal experience of online harassment. Full results for all robustness checks are provided in the Supplementary Information on OSF (https://osf.io/hnzba/files/nwvm3?view_only=defc9f54efc04e5a811a58bae22a1b5c).

In a final set of exploratory analyses, we examined participants' preferences when directly comparing support providers. Overall, 68% of participants selected LLM responses over human ones; this number rose to 81% in severe harassment cases, and LLM responses were rated as significantly more helpful overall (M = 3.80, SD = 1.01) compared to both human expert (M = 2.64, SD = 1.28, $t(860)$ = 14.83, 95% CI [1.01, 1.32], $d$ = 1.01, $p$ < .001) and non-expert responses (M =3.03, SD = 1.21, $t(860)$ = 10.13, 95%

CI [0.62, 0.92], $d = 0.62$, $p < .001$). We also examined participants' open-ended justifications and found that they valued LLMs' ability to signal empathy and provide practical guidance. Respondents frequently cited the presence of multiple actionable strategies, emotional validation, and a calm tone as central to their preference for LLM support.

These results not only validate the patterns observed in Study 1 but also show how linguistic empathy features work together to enhance coping outcomes for support recipients. Together, the findings indicate that LLM-generated support can meaningfully improve psychological well-being, both by fostering greater perceptions of being heard and by providing actionable guidance.

**Discussion**

As conversational AI becomes a more common source of emotional support, understanding how these systems influence psychological well-being is critical[53]. While our studies focus on general-purpose LLMs such as ChatGPT, our findings also speak to the growing class of AI companions designed explicitly for ongoing, personalized interaction[12]. Across two studies, we show that even without fine-tuning for emotional support, general-purpose LLMs can enhance individuals' coping self-efficacy in response to verbal harassment. Compared to non-expert humans and trained professionals, LLMs consistently produced language exhibiting stronger signals of perspective-taking, emotional validation, and action orientation. These linguistic markers made recipients feel more genuinely heard and increased their belief in their ability to cope with the harassment. Effects were strongest in more severe situations, where support is often more needed.

Our work makes three important contributions. First, while prior work has shown that LLM responses can receive higher empathy ratings than human responses[15,54], the present study extends this evidence by isolating three theoretically-grounded linguistic signals of empathy and demonstrating downstream effects on support recipients' outcomes, including feeling heard and coping self-efficacy. This integration of linguistic and behavioral evidence clarifies why and how AI-generated language can be experienced as empathic. Verbal harassment is a form of harm that operates through language by seeking to silence and discredit people. In this context, feeling heard is not only emotionally restorative but also reestablishes a sense of agency and control over the situation. When individuals feel heard, they regain confidence in their ability to respond and take constructive action, which strengthens coping self-efficacy.

Second, our findings suggest important contextual dynamics. In more severe harassment cases, where the emotional toll is greater and human supporters may disengage, LLM-generated responses maintained emotionally attuned and action-oriented language. Dedicated AI companions could leverage this stability in sustained relationships, offering more personalization and continuity of care over time. At the same time, because we prioritized controlled, one-off responses for our systematic comparison, our design limited our ability to examine how support unfolds across extended interactions. Future work should explore whether these benefits extend to longer-term exchanges and whether the observed psychological effects are sustained over time.

Third, our results also speak to broader questions of AI adoption. A common psychological barrier to adopting AI is the perception that it is inherently cold, emotionally limited, and ill-suited for subjective, emotion-based tasks[55,56]. The fact that participants in our studies consistently reported feeling more heard by LLMs than by human supporters suggests that this perception may be shifting. As people gain more direct experiences with AI systems that show empathic qualities, attitudes toward their suitability for emotionally supportive roles will likely increase.

Most importantly, we do not suggest replacing human mental health professionals. Rather, our findings highlight an opportunity to integrate AI into existing support systems. For example, LLMs can help untrained human supporters through training and language suggestions or serve as accessible first-line support when professional or peer help is unavailable. At the same time, human expertise remains indispensable for guiding AI toward deeper contextual understanding, ethical sensitivity, and appropriate care across diverse situations and cultures[57].

Taken together, our framework offers a novel lens for evaluating what makes supportive language effective, whether it comes from humans, general-purpose LLMs, or AI companions. As more human interactions are mediated through technology, the experience of feeling heard will increasingly depend on how language is used in these contexts. Our findings suggest that empathic AI can not only alleviate distress but also counteract the silencing effects of verbal harassment by restoring a sense of feeling heard and agency. By linking specific linguistic features to psychological outcomes, our findings demonstrate how empathy operates through language and provide a foundation for designing AI that meaningfully supports human well-being.

## Materials and Methods

### Stimuli and pretest

We first tested a set of 50 harassing online comments (Figure 8), selected from a publicly available dataset of social media comments previously coded as online harassment[58]. We employed a structured selection process to identify 50 comments that: 1) represented distinct categories of verbal harassment, 2) varied in apparent severity from mild to severe, and 3) maintained comparable linguistic features including length, complexity, and grammatical structure. With this controlled selection process, we aimed to ensure that experimental stimuli represent the spectrum of real-world harassment comments while minimizing confounding linguistic variables.

To assess harassment severity, we prompted GPT-4o to provide a severity rating on a 7-point scale (1 = "not at all serious" to 7 = "extremely serious")[59] and a categorical classification into one of three predefined classes (mild, moderate, severe)[60]. To enhance reliability and account for potential selection issues, each prompt was sent three times to GPT-4o (see Chen et al., 2023[61] for a similar approach and https://osf.io/2n9fv?view_only=defc9f54efc04e5a811a58bae22a1b5c for the full prompt skeleton).

Figure 8 here

Figure 8 lists all 50 statements and GPT-4o's severity rating and classification results. The model's quantitative ratings demonstrate significant variance across comments (M = 3.73, SD = 1.46), suggesting meaningful differentiation of severity levels. Ratings and classification were strongly convergent ($\eta^2$ = 0.8), with mean severity ratings significantly differing across classified categories in the expected direction (mild: M = 2.66, SD = 0.62; moderate: M = 4.51, SD = 0.77; severe: M = 7.00, SD = 0.00; $F(2, 147) = 290.29$, $p < .001$). Based on these empirically derived categories and additional criteria, including comment length and linguistic features, we selected six comments (three mild and three severe) for use in our main studies (Table 2). This systematic selection process ensures both high ecological validity and a tight experimental control.

Table 2 here

To further validate these severity assessments, we conducted a post-hoc check during Study 1. Specifically, the 604 human non-expert participants recruited through Prolific Academic ($M_{age}$ = 42.97,

SD = 12.91; 50.33% female) rated the severity of the harassment scenario they received using the same criteria as GPT-4o. These ratings showed a high degree of agreement with GPT-4o's pretest scores (ICC = 0.97, 95% CI [0.95, 0.98]), offering strong evidence that the model's severity classifications align closely with human judgments.

## Study 1: Comparative analysis of empathic listening features

### Participants and procedure

This study included N = 3,652 supportive responses generated by three groups of providers. We preregistered separate analysis plans for the human non-expert and expert samples, as they represented distinct populations of support providers and we aimed to examine the language each group produced individually before comparing them. For the LLM-generated responses, we did not preregister a separate analysis plan but documented all model versions, prompts, and sampling protocols on OSF to ensure transparency and reproducibility. Exclusion criteria were applied only to human participants, as preregistered; all LLM outputs were retained.

1. Human non-experts: 604 participants recruited through Prolific Academic ($M_{age}$ = 42.97, SD = 12.91; 50.33% female, 1.66% non-binary; https://aspredicted.org/zjr2-3z8z.pdf). Participants were instructed to imagine a friend receiving the harassing comment and provide supportive advice through an open text field. Each participant responded to one randomly drawn harassing comment selected in the pretest, resulting in a total of 604 responses.
2. Human experts: 216 mental health professionals recruited through a specialized recruitment agency ($M_{age}$ = 46.57, SD =11.62; 40.28% female, 0.93% non-binary; https://aspredicted.org/xw7m-h474.pdf), including licensed counselors, clinical psychologists, and psychiatrists (mean years of experience = 14.67, SD = 6.04), who were instructed to respond as if addressing a patient seeking guidance about the harassing comment in an open text field. Each expert responded to three comments randomly drawn from the six comments selected during the pretest. This resulted in a total of 648 responses, with a balanced distribution across the six comments.

3. LLMs: Four state-of-the-art LLMs, including GPT-4o (gpt-4o-2024-08-06), Claude Sonnet 3.5 (claude-3-5-sonnet-20240620), Llama 3.1 405B, and Pixtral 12B, were prompted using the same scenarios and instructions. The models were accessed through the official messages API endpoint of the OpenAI API (GPT-4o), Anthropic API (Claude Sonnet 3.5), Mistral API (Pixtral 12B), and Llama 3.1 405B was accessed through Microsoft Azure. We selected these models because they are widely used, high-performing, and accessible LLMs at the time of conducting this study (Hugging Face, 2024). We varied the gender and the age of the person seeking support to increase between-group scenario comparability. We collected ten responses per scenario from each model to account for potential variation in LLM responses (see https://osf.io/pxj4n?view_only=defc9f54efc04e5a811a58bae22a1b5c for the full prompt skeleton).

Sample sizes were chosen to balance comparability across provider groups and ensure sufficient power. Human targets were preregistered and reflect practical recruitment limits for experts. For LLMs, we generated multiple responses per scenario from four leading models to capture natural output variability. Because model-level patterns were highly consistent, we report aggregate results; detailed comparisons for each model are available on our OSF repository (https://osf.io/hnzba/files/e9d6t?view_only=defc9f54efc04e5a811a58bae22a1b5c).

All human and LLM responses addressed one of the six harassment scenarios (Table 2). We collected additional participant data (e.g., prior experience with harassment, social media use, demographics) as covariates. All participants provided informed consent prior to their participation. The study was approved by the University of St. Gallen Institutional Review Board (HSGEC-20231218). All methods were performed in accordance with regulations of the University of St. Gallen and the Declaration of Helsinki.

**Empathy feature ratings and content analysis**

For each response, we used GPT-4o to rate the three empathic listening features on a 7-point Likert scale (1 = very low, 7 = very high) and extract construct-consistent linguistic features:

- Perspective-taking: We prompted GPT-4o to rate the degree of perspective-taking and extract cognitive process words, hypothetical language, and other indicators of understanding the recipient's viewpoint. This operationalization aligns with established conceptualizations of perspective-taking in supportive communication[37].
- Emotional validation: We prompted GPT-4o to rate the level of emotional validation and extract emotional validation and affective language. This operationalization is consistent with measures of emotional validation[62].
- Action orientation: We prompted GPT-4o to rate the level of coping support and extract language suggesting specific actions aimed at protecting or supporting the recipient's well-being, consistent with frameworks used to evaluate coping support in therapeutic contexts.

The full prompt skeleton and extracted features can be found on OSF (https://osf.io/2hgvp?view_only=defc9f54efc04e5a811a58bae22a1b5c).

**Linguistic analysis (LIWC)**

We used a dictionary-based approach with the Linguistic Inquiry and Word Count[46] to examine the language structure of the supportive responses. LIWC evaluates language patterns by calculating word frequencies across various psychological and structural dimensions. Specifically, LIWC includes several summary variables designed to capture psychological constructs through algorithmic combinations of linguistic features[63]. For perspective-taking, we analyzed Analytical Thinking scores, which capture the use of cognitive processing and logical reasoning in language[46]. Higher analytical thinking scores are characterized by increased use of articles and prepositions, which suggests deeper cognitive engagement with the situation[64]. For emotional validation, we examined LIWC's Emotional Tone and Authenticity scores. Emotional tone quantifies the valence of emotional language, with higher scores indicating more positive emotional expressions[65], while Authenticity measures the degree of emotionally open language by looking at first-person singular pronouns (e.g., I, me), present-tense verbs (e.g., am, feel), emotionally expressive language, and a more conversational style[47]. Finally, we used LIWC's summary measure of Clout as a proxy to evaluate more action-oriented responses. Kacewicz et al. (2014) showed that higher Clout scores reflect language conveying guidance and confidence, which aligns with our conceptualization of action orientation as communicating agency and the ability to guide others[48].

**Coping strategy classification**

To analyze coping strategies, we used a combination of LLM and rule-based classification. First, we used GPT-4o to extract all distinct coping strategies mentioned in each support response. Specifically, we prompted the model to parse the response, identify discrete strategies, and return results in a structured JSON format. Next, we applied a keyword-based taxonomy to categorize each strategy as either approach-oriented or avoidance-oriented. We developed the taxonomy iteratively based on established distinctions in coping research[42] and incorporating patterns we observed in the responses. We first tagged each strategy with a specific category (e.g., ignoring, blocking, reporting, seeking social support, using humor), based on a comprehensive set of regular expression rules designed to detect language patterns across responses. Then we mapped these categories as either approach-oriented strategies, which directly engage with the stressor or build coping capacity (e.g., confronting or reporting the harasser, documenting evidence, seeking social support, learning new skills) or as avoidance-oriented strategies, which create emotional or physical distance from the stressor (e.g., ignoring, blocking, taking breaks, adjusting privacy settings, disengaging emotionally). Our final taxonomy covered over 95% of all extracted strategies, with residual cases labeled as "Other". The detailed taxonomy can be found on OSF (https://osf.io/v87te?view_only=defc9f54efc04e5a811a58bae22a1b5c).

**Analytical approach**

We compared the ratings of our three empathy features (perspective-taking, emotional validation, and action orientation) across provider types using Welch's t-tests for each pairwise contrast, with Holm–Bonferroni adjustment for multiple comparisons (provider contrasts; severity contrasts). Effect sizes are reported as Cohen's d (unpooled SDs) with 95% confidence intervals; two-tailed $\alpha = .05$. To test robustness, we repeated all comparisons using non-parametric Wilcoxon tests. Conclusions were identical. LIWC and coping strategy data were compared using t-tests and chi-square tests. All analyses were conducted in Python and R (v4.3).

**Study 2: Recipient perceptions and psychological outcomes**

**Participants and procedure**

In this preregistered study (https://aspredicted.org/ghw2-tsff.pdf) we recruited a new panel of participants through Prolific Academic (N = 431; $M_{age}$ = 36.14, SD = 13.6; 50.58% female) to evaluate a representative subsample of support responses from Study 1. To ensure both experimental control and high ecological validity, we implemented multi-level randomization. First, participants were randomly assigned to one of the six online harassment cases. They were told to "Imagine the following situation: Someone wrote '[comment]' on social media to you. You do not know who wrote it and you are unsure how to respond." After 20 seconds, participants were told, "You reached out for support and received the following response: [response]."

We pre-selected representative responses from each provider type (GPT-4o, human non-experts, and mental health professionals) for each harassment scenario from Study 1, to reflect typical word count and empathic listening scores for that provider. We randomly assigned each participant to one response per provider type. Each participant evaluated all three support responses (one from each provider) for a specific harassment scenario, with the order of presentation randomized. This design produced 54 unique responses (6 scenarios × 3 provider types × 3 responses), resulting in a total of 1,296 evaluations.

**Measures**

After reading each scenario case and the corresponding response, participants rated the response that they received across multiple dimensions using 7-point Likert-type scales (1 = strongly disagree, 7 = strongly agree). First, participants rated the responses on the extent it made them feel heard, using a validated three-item scale developed by Roos et al. (2023; "I feel heard," "I feel listened to," "I feel understood"; $\alpha = .96$)[21]. Next, participants evaluated their perceived coping self-efficacy on four items adapted from Chesney et al., (2006; e.g., "This response helps me to break an upsetting problem down into smaller parts," "This response helps me to make a plan of action and follow it"; $\alpha = .88$)[51]. Participants also rated each response on emotional validation (four items adapted from the IPIP Empathy scale e.g., "Feels my emotions,"; $\alpha = .93$)[66], perspective-taking (four items from the Interpersonal Reactivity Index, e.g., "Tries to look at my side of the situation"; $\alpha$ = .82)[67], and action-oriented (two custom-developed items, e.g., "Recommends taking action to resolve the issue"; $\alpha$ = .85).

We included exploratory measures for psychological well-being (adapted from the Warwick-Edinburgh Mental Well-Being Scale, e.g., "The response makes me more relaxed;" $\alpha$ = .95)[68]. Participants also rated the perceived helpfulness of the response on a 5-point scale from "extremely helpful" to "not at all helpful." Finally, participants were shown all three responses they had evaluated side-by-side and asked to select the one they preferred.

We also collected additional participant data as covariates, including prior experience with online harassment (e.g., "Have you ever received a comment online that was accusing, threatening, provocative, or rude?"), perceived social support[69] and demographics (age, gender, ethnicity, education). These covariates were used to check for differences and ensure consistency of effects.

All participants provided informed consent prior to their participation. The study was approved by the University of St. Gallen Institutional Review Board (HSGEC-20231218). All methods were performed in accordance with regulations of the University of St. Gallen and the Declaration of Helsinki.

**Analytical approach**

We used one-way ANOVAs to compare the effects of provider type on outcome variables, with harassment severity as a between-subjects factor. We used planned pairwise contrasts to compare all three provider types (GPT-4o, human non-experts, and human experts). All p-values were adjusted for multiple comparisons using the Holm-Bonferroni method, and effect sizes are reported as Cohen's d. We estimated a moderated mediation model using the lavaan package in R to test whether differences in provider type predicted coping self-efficacy via feeling heard, and whether this indirect effect was moderated by harassment severity. We implemented orthogonal contrast coding to compare (1) GPT responses versus human responses and (2) experts versus non-experts. To test for potential moderation, we included interaction terms between these contrasts and severity. We used 10,000 bootstrap samples to estimate indirect effects and 95% confidence intervals. Robustness checks included re-running all analyses, controlling for response length, as well as subgroup analyses among participants with personal harassment experience. All analyses were conducted using Python and R (v4.3), with $\alpha = .05$.

**Author Contributions**

A.B. designed the research; P.W. collected the data; A.B. and P.W. analyzed different components of the data; A.B. wrote the paper with input from P.W. and C.H.; C.H. contributed to conceptualization, study design, and interpretation of findings.

**Competing Interest Statement**

The authors declare no competing interests.

**Data and code availability**

All data and code are publicly available at: https://osf.io/hnzba/files?view_only=defc9f54efc04e5a811a58bae22a1b5c.We use GitHub to ensure proper versioning, while linking it to OSF provides traceability, transparency, and view-only access for peer review.

**Funding**

This research was funded by the University of St. Gallen, including support from the Green Box Project for Research Innovation.

## Figures and Tables

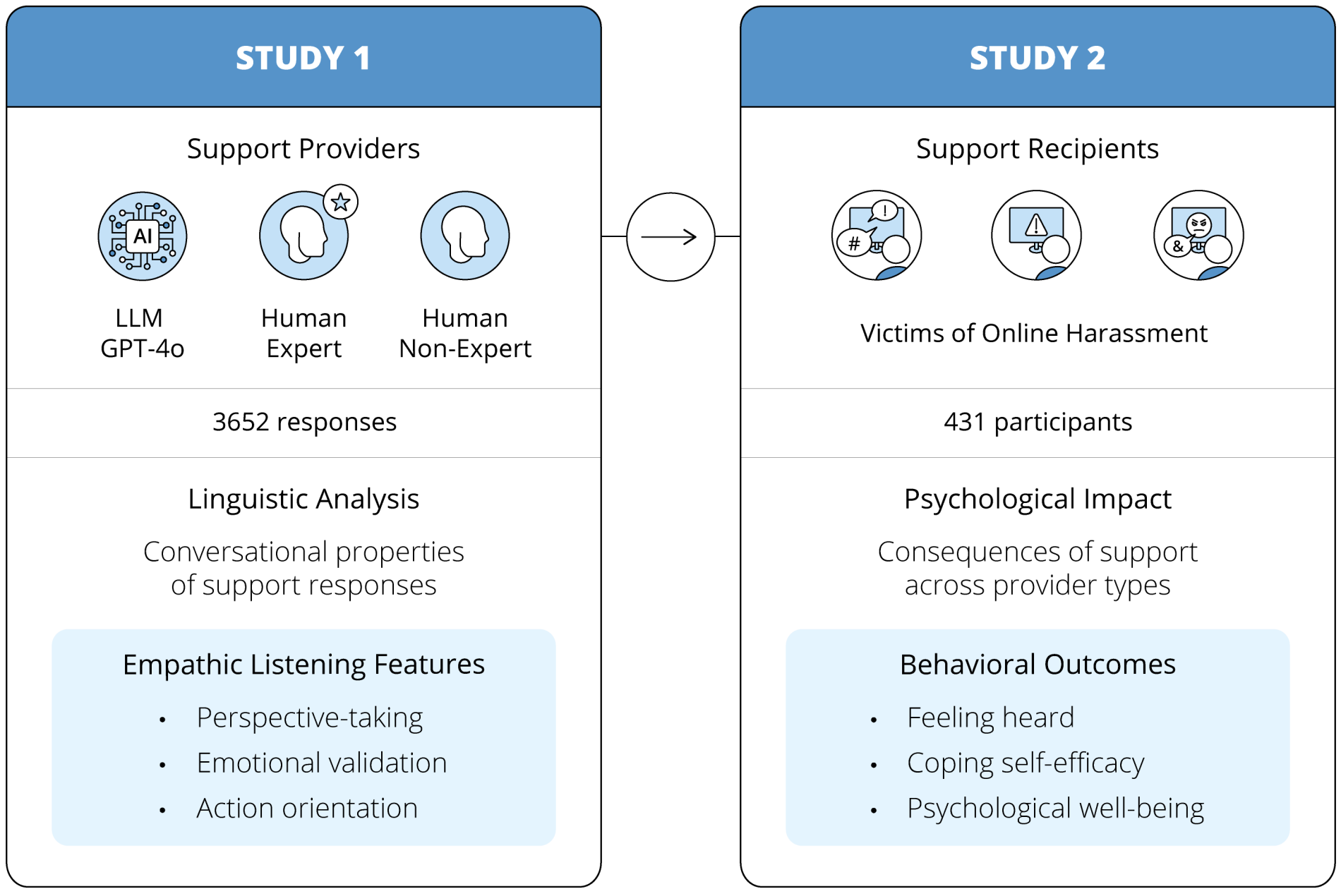

**Figure 1.** Research framework with a dual-sided approach to studying coping support.

Study 1 focuses on the support provider side, analyzing how linguistic signals of empathy are encoded in supportive messages. Study 2 focuses on the support recipient side, examining how these same messages are perceived by victims of online harassment

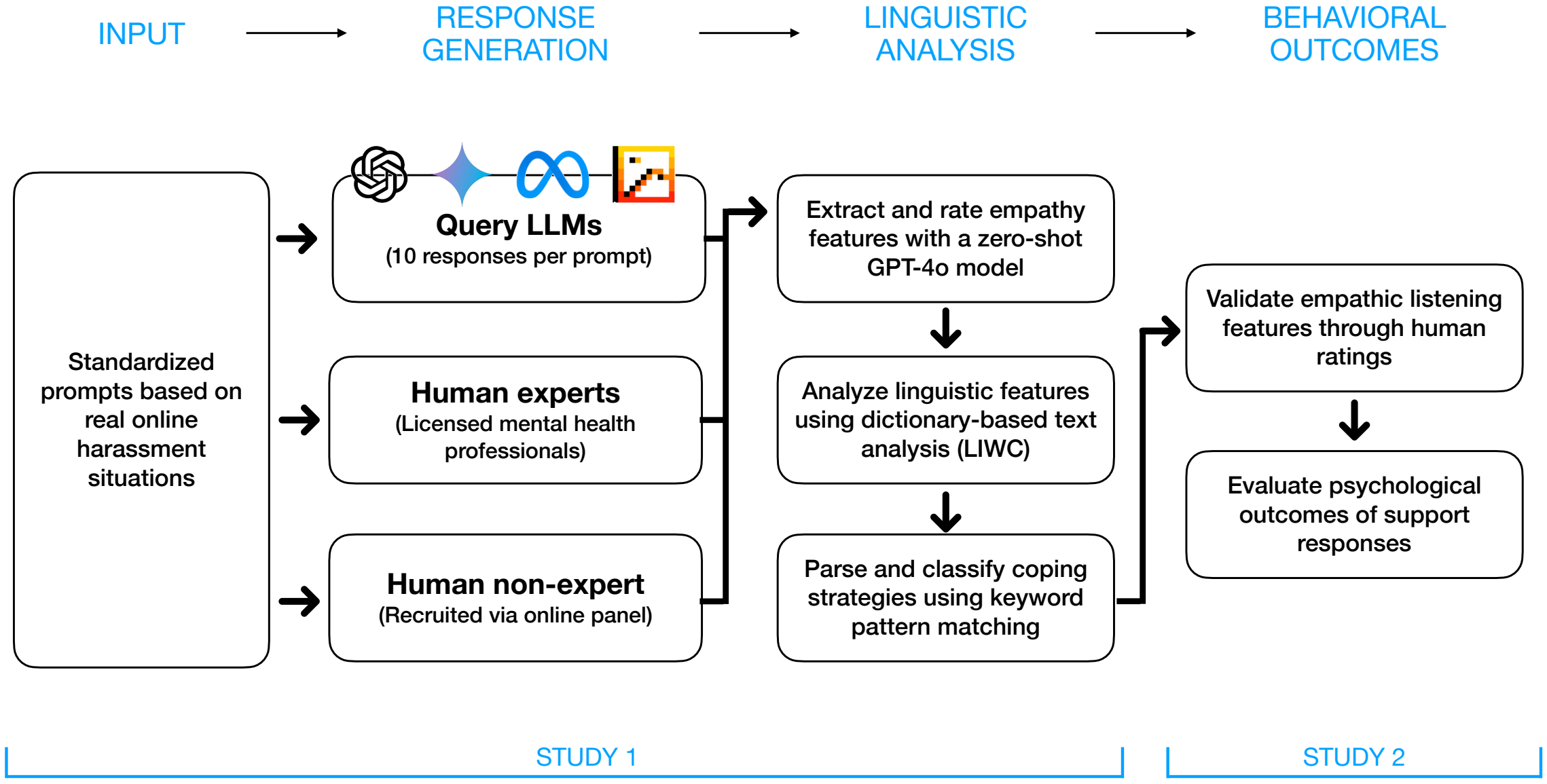

**Figure 2.** Overview of experimental pipeline.

Standardized verbal harassment scenarios were used to elicit supportive responses from LLMs, licensed mental health professionals, and human non-experts (Study 1). Responses were analyzed for linguistic empathy features, linguistic style, and coping strategies, then presented to participants exposed to the same verbal harassment scenarios to validate features and assess psychological outcomes (Study 2).

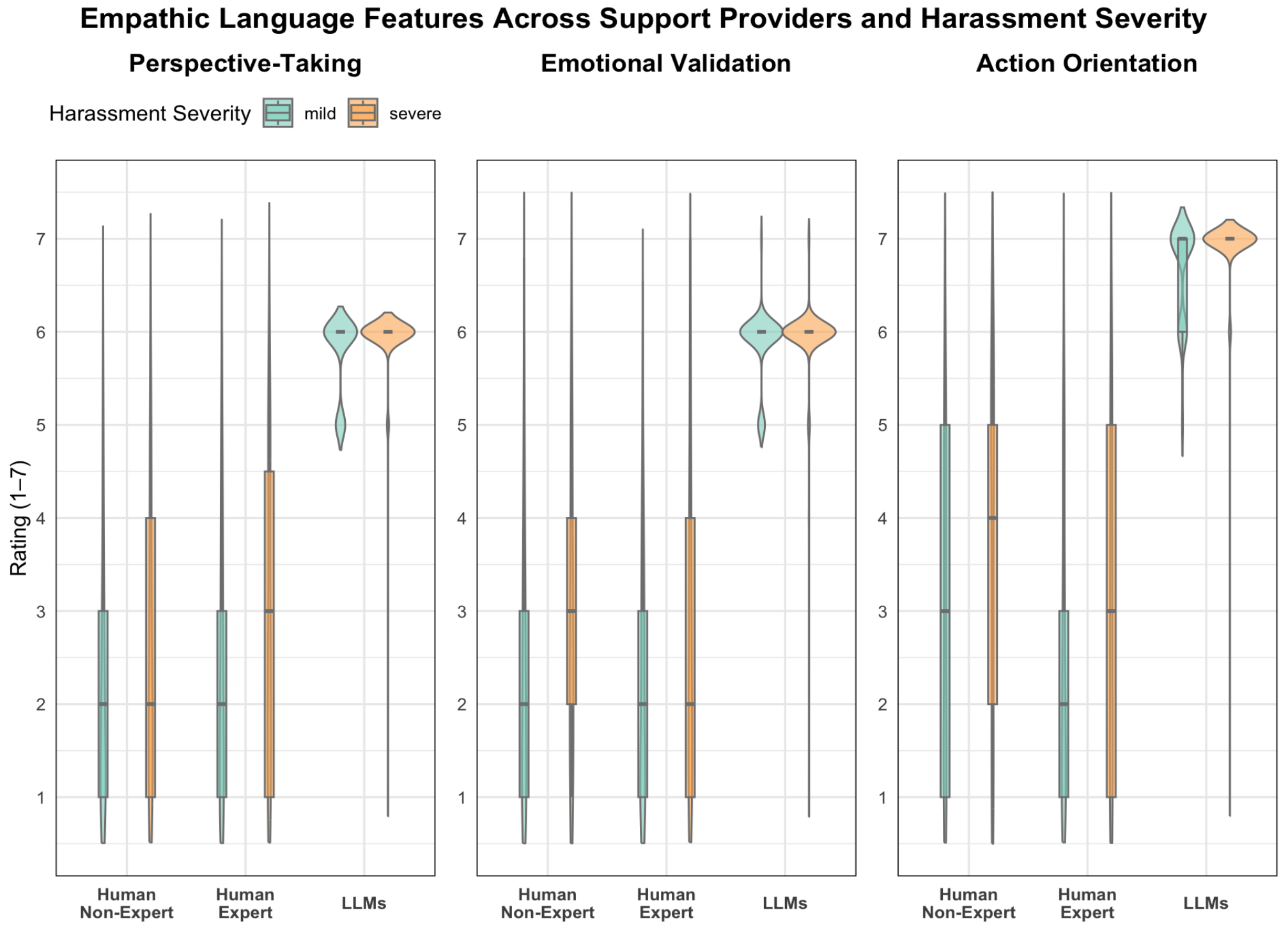

**Figure 3.** Empathic listening features across provider types and harassment severity.

Ratings (1–7) of perspective-taking, emotional validation, and action orientation were generated by GPT-4o for responses from human non-experts, human experts, and LLMs to mild (green) and severe (orange) verbal harassment scenarios. Violin plots show score distributions; horizontal lines indicate medians. LLMs scored significantly higher than humans across all features, and all groups showed higher ratings in severe scenarios.

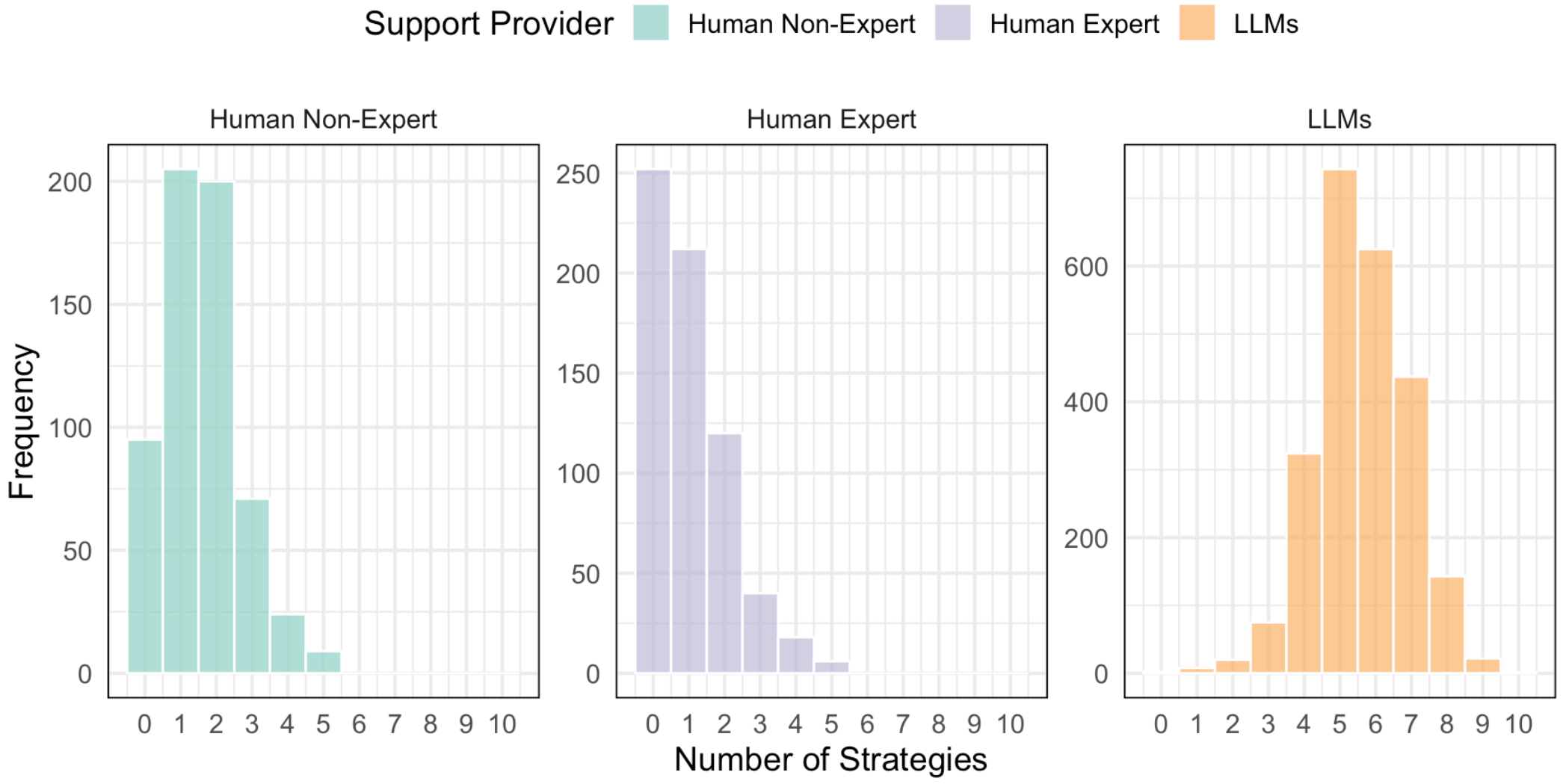

**Figure 4.** Number of coping strategies per response by support provider type. Distribution of distinct coping strategies suggested by human non-experts, human experts, and LLMs in Study 1. LLMs generated more strategies per response than either human group.

**Strategy type by Support Provider**

| | Human Non-Expert | Human Expert | LLMs |
|---|---|---|---|
| Distance | 32.3% | 23.4% | 3.8% |
| Ignore | 22.2% | 16.8% | 16.9% |
| Block | 18.4% | 9.1% | 12.8% |
| Confront | 9.5% | 7.6% | 17.3% |
| Seek Social Support | 5.8% | 13.1% | 13.1% |
| Report | 4.9% | 7.9% | 8.8% |
| Use Humor | 1% | 1.3% | 8.3% |
| Self-Care | 0.3% | 2.5% | 2.6% |
| Use Empathy | 0.2% | 3.3% | 1.5% |
| Set Boundaries | | 0.1% | 2.7% |
| Document | 1.3% | 0.4% | 1.9% |
| Ask Clarification | | 0.3% | 1.4% |
| Reframe | 0.3% | 0.7% | 1.5% |
| Educate | 0.1% | | 1.5% |
| Take High Road | | | 0.8% |
| Learn Skills | 0.3% | 0.9% | 1% |
| Limit Exposure | 0.4% | 1% | 0.7% |
| Express Feelings | | 1.3% | 0% |
| Use Positivity | 0.3% | | 0.7% |
| Self-Affirm | 0.3% | 0.7% | 0.3% |
| Redirect Conversation | | | 0.3% |
| Don't Take Personally | 0.1% | 0.1% | 0.6% |
| Take Breaks | 0.1% | 0.1% | 0.2% |

Blue text: Approach-oriented strategies | Purple text: Avoidance-oriented strategies

**Figure 5.** Distribution of coping strategy types by provider type. Relative frequency of distinct coping strategies suggested by human non-experts, human experts, and LLMs in Study 1. Approach-oriented strategies are shown in blue, avoidance-oriented strategies in purple. LLMs recommended a higher proportion of approach-focused strategies, while human non-experts relied predominantly on avoidance. Percentages may not sum to 100% because ~5% of strategies could not be classified.

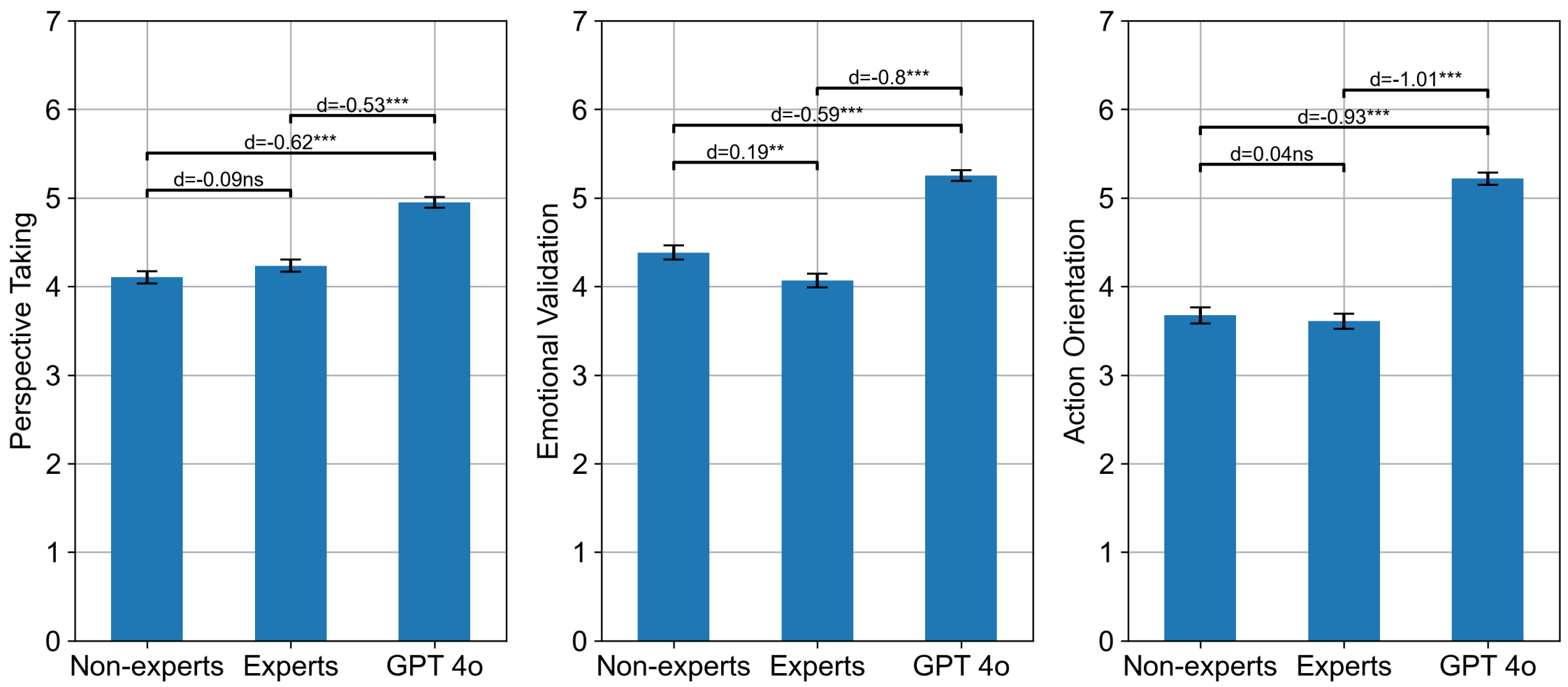

**Figure 6.** Empathic listening features rated across support providers by recipients.

Mean ratings of perspective-taking, emotional validation, and action orientation in responses from human non-experts, human experts, and GPT-4o. Effect sizes (d) show pairwise contrasts; *** = $p < .001$, ** = $p < .01$, ns = non-significant. LLM responses were rated higher than human responses on all three features.

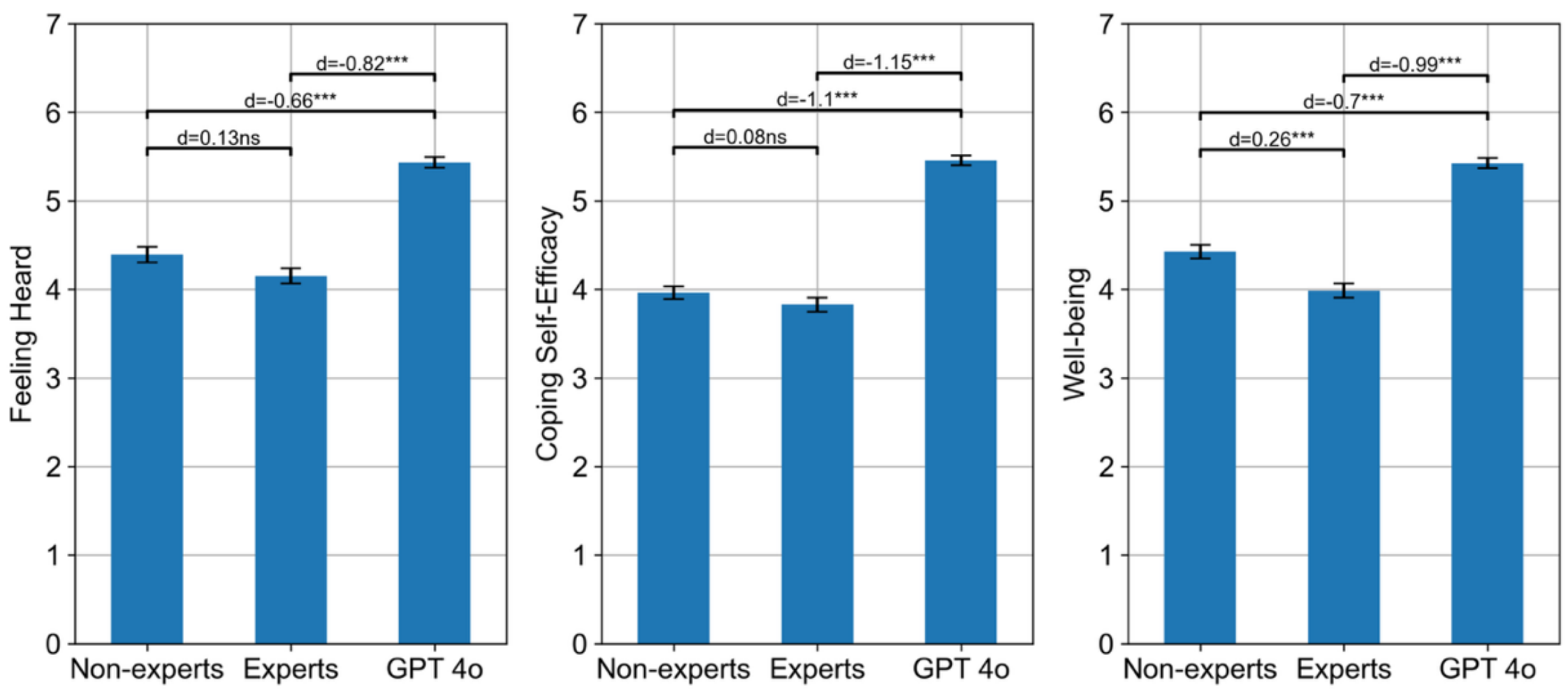

**Figure 7.** Psychological outcomes of supportive responses across provider types.

Mean ratings of feeling heard, coping self-efficacy, and overall well-being after reading responses from human non-experts, human experts, and GPT-4o. Effect sizes (d) show pairwise contrasts; *** = $p < .001$, ** = $p < .01$, ns = non-significant. LLM responses were associated with significantly higher ratings on all three outcomes compared with human responses.

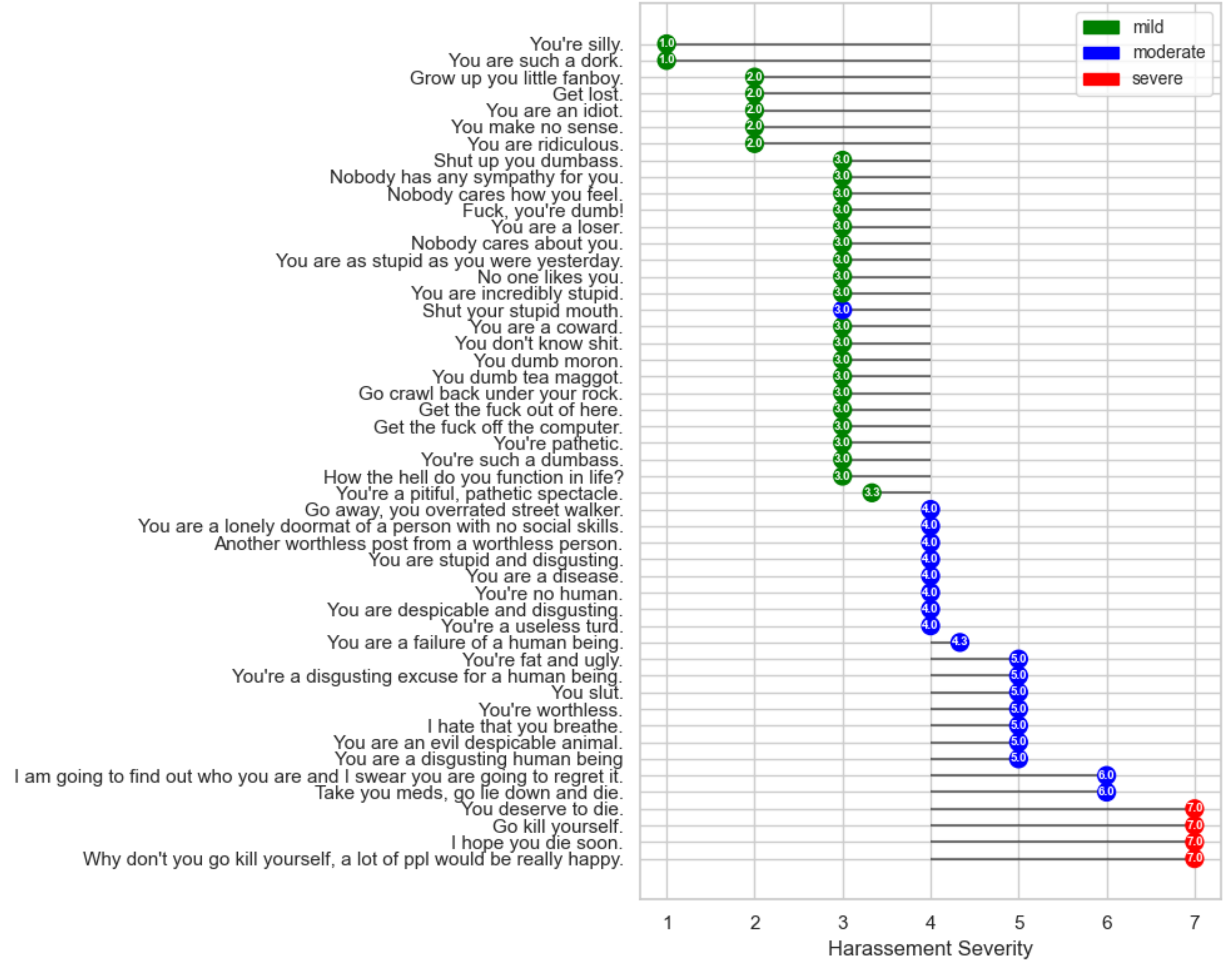

**Figure 8.** Severity ratings and classifications of 50 harassment comments by GPT-4o.

Mean severity scores (1 = not at all serious, 7 = extremely serious) and categorical classifications (mild, moderate, severe) for each comment in the pretest. Ratings were averaged across three independent prompts per comment.

**Table 1.** Examples of supportive responses by provider type and harassment severity.

| Severity | Support provider | Exemplary response |
|---|---|---|
| Mild | Human non-expert | I would say you have the choice to say nothing and block them. Or you could just send them positive vibes and state that maybe they are having a bad day. You can feed into that negativity or do better than them. |
| | Human expert | We can't change other people—how they think or what they may say. We can only change and modify how we respond. If someone is being mean in this manner, you must decide for yourself how best to approach this. What will make you feel better at the end of the day? If you respond, how do you know the person will even see or care about your response. You can block negative people. You can choose not to post if such responses upset you. |
| | LLM (GPT-4o) | I'm sorry to hear that someone left such an unkind comment. It's important to remember that you don't have to engage with negativity. However, if you choose to respond, here's a way to do it kindly and maturely: "Thank you for sharing your thoughts. I believe in keeping our interactions positive and respectful. Wishing you a wonderful day!" It's often best to take the high road and not let negative comments affect your peace of mind. |
| Severe | Human non-expert | Oh that is so awful. Don't pay any attention to social media posts. People are just mean and hateful online. They hide behind their keyboard because they think they are better than everyone else. I am always here for you whenever you need someone to talk to. You are loved. |
| | Human expert | First I'm sorry you had to read that, but know that you're more valuable than that person knows. Please never think someone else's negative opinion or comment towards you has any weight. This world wouldn't be the same without you. Let us process together how this comment has made you feel and what it has brought up for you. I advise disabling that person's access to your account and reporting them to the social media websites for inappropriate use. |
| | LLM (GPT-4o) | I'm sorry to hear that you're experiencing this negativity online. Facing such messages can be really upsetting. Here's a possible response you might consider: "Your comment is very hurtful and inappropriate. Let's aim to support and uplift one another instead of spreading negativity. Wishing you kindness and understanding." It's also important to prioritize your well-being. Consider reporting and blocking the person who made the comment if you feel unsafe, and reach out to friends, family, or mental health professionals for support. Remember, you're not alone, and there are people who care about you. |

**Table 2.** Pretested online harassment comments selected for the main studies.

| Harassment Severity | Harassing Comment |
|---|---|
| **Mild** | You are a loser. |
| | You are an idiot. |
| | Get lost. |
| **Severe** | Go kill yourself. |
| | I hope you die soon. |
| | You deserve to die. |